\documentclass[10pt, twocolumn, twoside]{IEEEtran}

\usepackage{cite}
\usepackage{enumerate}
\usepackage{graphicx}
\usepackage[cmex10]{amsmath} 
\usepackage{amssymb}
\usepackage{mathtools}
\usepackage{xspace}
\usepackage{subfigure}
\usepackage[ruled,vlined]{algorithm2e} 
\usepackage{colonequals}
\usepackage[ mathscr]{euscript}
\usepackage{fixltx2e}    

\usepackage{epsfig}
\usepackage{epstopdf}

\usepackage{tikz}
\usepackage{pgfplots}

\usetikzlibrary{arrows,shapes,backgrounds,plotmarks,positioning}

\pgfplotsset{compat=newest}                         
\pgfplotsset{plot coordinates/math parser=false}
\newlength\figureheight
\newlength\figurewidth

\newtheorem{theorem}{Theorem}[section]

\newtheorem{proposition}[theorem]{Proposition}

\newtheorem{example}[theorem]{Example}

\newcommand{\C}{\mathcal{K}}         
\newcommand{\Pak}{\mathcal{P}}       
\newcommand{\pv}{\mathbf{p}}         
\newcommand{\F}{\mathbb{F}}          
\newcommand{\Has}{\mathcal{H}}       
\newcommand{\rv}{\mathbf{r}}         
\newcommand{\R}{\mathcal{R}}         
\newcommand{\Ru}{\R_{\alpha}}         
\newcommand{\RuPS}{\hat{\R}_{\alpha}}         

\newcommand{\X}{\mathcal{X}}         
\newcommand{\Y}{\mathcal{Y}}         

\newcommand{\Iset}{\mathcal{I}}
\newcommand{\W}{\mathcal{W}}

\newcommand{\Pat}{\{\X_i\}_{i\in\Iset}}

\newcommand{\N}{\mathbb{N}_0^K}

\newcommand{\Real}{\mathbb{R}_{+}}

 {\begin{list}{}%
         {\setlength{\leftmargin}{#1}}%
         \item[]%
 }
 {\end{list}}

\begin{document}

\title{Estimating Minimum Sum-rate for Cooperative Data Exchange}


\author{
\IEEEauthorblockN{Ni~Ding\IEEEauthorrefmark{1}, Rodney~A.~Kennedy\IEEEauthorrefmark{1} and Parastoo~Sadeghi\IEEEauthorrefmark{1}}\\
\IEEEauthorblockA{\IEEEauthorrefmark{1}The Research School of Engineering, College of Engineering and Computer Science, the Australian National University (ANU), Canberra, ACT 2601\\
Email: $\{$ni.ding, rodney.kennedy, parastoo.sadeghi$\}$@anu.edu.au}
}


\maketitle

\begin{abstract}

This paper considers how to accurately estimate the minimum sum-rate so as to reduce the complexity of solving cooperative data exchange (CDE) problems. The CDE system contains a number of geographically close clients who send packets to help the others recover an entire packet set. The minimum sum-rate is the minimum value of total number of transmissions that achieves universal recovery (the situation when all the clients recover the whole packet set). Based on a necessary and sufficient condition for a supermodular base polyhedron to be nonempty, we show that the minimum sum-rate for a CDE system can be determined by a maximization over all possible partitions of the client set. Due to the high complexity of solving this maximization problem, we propose a deterministic algorithm to approximate a lower bound on the minimum sum-rate. We show by experiments that this lower bound is much tighter than those lower bounds derived in the existing literature. We also show that the deterministic algorithm prevents from repetitively running the existing algorithms for solving CDE problems so that the overall complexity can be reduced accordingly.
\end{abstract}


\section{introduction}
\label{sec:Intro}

Due to the growing amount of data exchange over wireless networks and increasing number of mobile clients, the base-station-to-peer (B2P) links are severely overloaded. It is called the `last mile' bottleneck problem in wireless transmissions. Cooperative peer-to-peer (P2P) communications is proposed for solving this problem. The idea is to allow mobile clients to exchange information with each other through P2P links instead of solely relying on the B2P transmissions. If the clients are geographically close to each other, the P2P transmissions could be more reliable and faster than B2P ones.

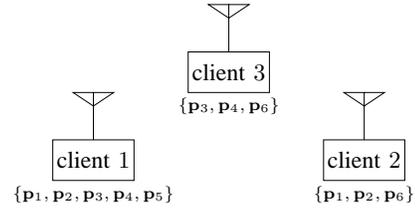
\begin{figure}[tpb]
	\centering
    \scalebox{0.9}{\begin{tikzpicture}

\draw (-2.6,0.3) rectangle (-1.4,-0.3);
\node at (-2,0) {client $1$};
\draw (-2,0.3)--(-2,1)--(-1.7,1)--(-2,0.8)--(-2.3,1)--(-2,1);
\node at (-2,-0.5) {\scriptsize $\{\pv_1,\pv_2,\pv_3,\pv_4,\pv_5\}$};

\draw (2.6,0.3) rectangle (1.4,-0.3);
\node at (2,0) {client $2$};
\draw (2,0.3)--(2,1)--(1.7,1)--(2,0.8)--(2.3,1)--(2,1);
\node at (2,-0.5) {\scriptsize $\{\pv_1,\pv_2,\pv_6\}$};

\draw (-0.6,1.6) rectangle (0.6,1);
\node at (0,1.3){client $3$};
\draw (0,1.6)--(0,2.3)--(0.3,2.3)--(0,2.1)--(-0.3,2.3)--(0,2.3);
\node at (0,0.8) {\scriptsize $\{\pv_3,\pv_4,\pv_6\}$};

\node at (0,2.5) {$\phantom{a}$};

\end{tikzpicture}}
	\caption{An example of CDE system: There are three clients that want to obtain six packets. The has-sets are $\Has_1= \{\pv_1,\pv_2,\pv_3,\pv_4,\pv_5\}$, $\Has_2= \{\pv_1,\pv_2,\pv_6\}$ and $\Has_3=\{\pv_3,\pv_4,\pv_6\}$.}
	\label{fig:CDESystem}
\end{figure}

Consider the situation when a base station wants to deliver a set of packets to a group of clients. Denote $\Pak=\{\pv_1,\dotsc,\pv_L\}$ the packet set and $\C=\{1,\dotsc,K\}$ the client set. Due to the fading effects of wireless channels, after several broadcasts via B2P links, there may still exist some clients that do not obtain all the packets. Fig.~\ref{fig:CDESystem} shows an example when the base station wants to disseminate $6$ packets to $3$ clients. In this figure, $\Has_j$ is the has-set that denotes the packets obtained by client $j$ after B2P transmissions. Since the clients' knowledge of the packet set may be complementary to each other, instead of relying on retransmissions from the base station, the clients can broadcast combinations of the packets they know via P2P wireless links so as to help the others recover the missing packets. We call Fig.~\ref{fig:CDESystem} cooperative data exchange (CDE) system. For this kind of systems, there is a so-called CDE problem: how to find an efficient transmission strategy that achieves the \textit{universal recovery} (the situation when all clients recover the entire packet set).

Let $\rv=(r_1,\dotsc,r_K)$ be a transmission strategy, where $r_j$ denotes the total number of linear combinations transmitted by client $j$. The CDE problem is usually expressed in the general form of \cite{CourtIT2014}
\begin{align} \label{eq:Problem}
    &\quad \min_{\rv} f(\rv) \nonumber \\
    &\text{s.t.} \sum_{j\in\X}r_j \geq |\bigcap_{j\in\C\setminus\X}\Has_j^c|, \forall{\X\subset\C}, \nonumber \\
    &\quad \sum_{j\in\C}r_j = \alpha,
\end{align}
where $\alpha$ denotes the transmission budget, the designated total number of transmissions among the clients. In \eqref{eq:Problem}, the expression of $f(\rv)$ is determined based on what kind of strategy is considered to be efficient. For example, if $f(\rv)=\mathbf{w}^\intercal\rv$, the most efficient strategy is the one that minimizes the weighted-sum of transmissions \cite{CourtIT2014}; if $f(\rv)=\sum_{j\in\C}r_j\log(r_j)$, the most efficient strategy is the one that distribute the transmission rate in the fairest way \cite{Milo2012}. For most of the algorithms that solve a particular CDE problem, the value of $\alpha$ is assumed to be known a priori. For example, the algorithms proposed in \cite{CourtIT2014,Milo2012} find minimum weighted sum-rate and fairest strategies, respectively, for a given value of $\alpha$.

However, problem \eqref{eq:Problem} does not have solutions for all values of $\alpha$. In fact, there exists a minimum sum-rate $\alpha^*$ so that the constraint set in \eqref{eq:Problem} is nonempty only if $\alpha\geq\alpha^*$. But, the value of $\alpha^*$ is usually unknown in advance. On the other hand, the algorithms proposed in \cite{CourtIT2014,Milo2012} can check the feasibility of a given value of $\alpha$ (i.e., whether there exists a strategy that can achieve universal recovery under budget $\alpha$). Therefore, for the minimization problem over the minimum sum-rate strategy set (problem \eqref{eq:Problem} when $\alpha=\alpha^*$), one can start with any arbitrary value of $\alpha$ and adjust it to the feasible values accordingly, e.g., setting $\alpha$ as the lower bound on $\alpha^*$ derived in \cite{Roua2010,SprintRand2010}, run the algorithms in \cite{CourtIT2014,Milo2012} and increase the value of $\alpha$ until a feasible solution is found. By doing so, the algorithms in \cite{CourtIT2014,Milo2012} should be called for many times until the final solution is found. But, the complexity of running algorithms proposed in \cite{CourtIT2014,Milo2012} is not low. These algorithms involve submodular function minimizations, the complexity of which is at least $O(K^5\gamma+K^6)$.\footnote{There are many algorithms proposed for submodular function minimization problem. To our knowledge, the algorithm proposed in \cite{Goemans1995} has the lowest complexity $O(K^5\gamma+K^6)$, where $\gamma$ is the complexity of evaluating a submodular function.} For this reason, estimating a value of minimum sum-rate that is close to $\alpha^*$ prevents from repetitively running algorithms proposed in \cite{CourtIT2014,Milo2012} and, therefore, helps reduce the overall complexity of solving problem \eqref{eq:Problem}.

In this paper, we study how to accurately estimate the minimum sum-rate $\alpha^*$ for both CDE systems that allow packet splitting (PS-CDE) and CDE systems that do not allow packet splitting (NPS-CDE). We start the study with an existing result derived in \cite{CourtIT2014,Milo2012}: the crossing submodularity of $|\bigcap_{j\in\C\setminus\X}\Has_j^c|$. By using a necessary and sufficient condition given in \cite{Fujishige2005} for the constraint set in \eqref{eq:Problem} to be nonempty, we show that the exact value of $\alpha^*$ can be determined by a maximization over all possible partitions of the client set $\C$, which is NP-hard \cite{Goldschmidt1994}. Instead of calculating the exact value of $\alpha^*$, we propose a deterministic algorithm to estimate a lower bound on $\alpha^*$. We show by experiments that the lower bound found by the deterministic algorithm is much tighter than the ones derived in \cite{Roua2010,SprintRand2010}. In fact, the experiment results show that the deterministic algorithm returns the exact value of minimum sum-rate in most cases. In this paper, we also discuss how our results are related to those in \cite{MiloDivConq2011,CourtIT2014,Roua2010,SprintRand2010} and prove that any PS-CDE problem can be converted to an NPS-CDE problem.

\section{System Model and Problem Statement}
\label{sec:system}

Let $\Pak=\{\pv_1,\dotsc,\pv_L\}$ be the packet set containing $L$ linearly independent packets. Each packet $\pv_i$ belongs to a field $\F_q$. The system contains $K$ geographically close clients. Define the client set as $\C=\{1,\dotsc,K\}$. Each client $j\in\C$ initially obtains $\Has_j\subset\Pak$. Here, $\Has_j$ is called the \textit{has-set} of client $j$. We also denote $\Has_j^{c}=\Pak\setminus\Has_j$ as the packet set that is missing at client $j$. The clients are assumed to collectively know the the packet set, i.e., $\cup_{j\in\C}\Has_j=\Pak$. The P2P wireless links between clients are error-free, i.e., any information broadcast by client $j$ can be heard losslessly by client $j'$ for all $j'\in\C\setminus{\{j\}}$. The clients broadcast linear combinations of the packets in their has-sets in order to help each other recover the entire packet set $\Pak$. For example, in the CDE system in Fig.~\ref{fig:CDESystem}, client $1$ broadcasting $\pv_1+\pv_3$ helps client $2$ recover $\pv_3$ and client $3$ recover $\pv_1$, and client $2$ broadcasting $\pv_1+\pv_6$ helps client $1$ recover $\pv_6$ and client $3$ recover $\pv_1$.

For a transmission strategy $\rv=(r_1,\dotsc,r_K)$, where $r_j$ denotes the total number of linear combinations transmitted by client $j$, we call $\sum_{j\in\C}r_j$ the sum-rate of strategy $\rv$. Denote $\Ru$ the set that contains all transmission strategies that achieve universal recovery and have sum-rate equal to constant $\alpha$. $\Ru$ is nonempty only if $\alpha$ is greater or equal to the minimum sum-rate $\alpha^*$. For example, consider the CDE system in Fig.~\ref{fig:CDESystem} and assume it is an NPS-CDE system. Since $\rv\in\N$, $\alpha\in\mathbb{N}_0$, i.e., we only consider nonnegative integer values of $\alpha$. In this system, $\alpha\leq{3}$, $\Ru=\emptyset$, i.e., there is no strategy with a sum-rate less than or equal to $3$ that can achieve universal recovery. But, if $\alpha=4$, there exist three strategies that achieve universal recovery: $\R_4=\{(3,0,1),(3,1,0),(2,1,1)\}$. For example, in strategy $(3,0,1)$, client $1$ broadcasting $\pv_1+\pv_3$, $\pv_2+\pv_4$ and $\pv_5$ and client $3$ broadcasting $\pv_6$ can achieve the universal recovery. Therefore, the minimum sum-rate is $\alpha^*=4$. Assume Fig.~\ref{fig:CDESystem} is a PS-CDE system. Since $\rv\in\Real^K$, $\alpha\in\Real$, i.e., we consider the nonnegative real values of $\alpha$. It can be shown that $\alpha^*=3.5$ with the corresponding strategy set being $\R_{3.5}=\{(2.5,0.5,0.5)\}$ and $\Ru=\emptyset$ if $\alpha<3.5$. Note, the way to implement strategy $(2.5,0.5,0.5)$ is to break each packet into two chunks, e.g., the has-set of client $3$ would be $\{\pv_3^{(1)},\pv_3^{(2)},\pv_4^{(1)},\pv_4^{(2)},\pv_6^{(1)},\pv_6^{(2)}\}$. Letting client $1$ transmit $5$ chunks and both client $2$ and client $3$ transmit $1$ chunk is sufficient to achieve the universal recovery. In this case, the packet normalized strategy is $(2.5,0.5,0.5)$.

It is shown in \cite{CourtIT2014} that the CDE problems can be formulated by \eqref{eq:Problem} in general. Note, \eqref{eq:Problem} is an integer programming problem for NPS-CDE systems. It can be seen from the above examples that for any CDE system there exist a minimum sum-rate $\alpha^*$ such that $\Ru$ is nonempty only if $\alpha\geq\alpha^*$, i.e., problem \eqref{eq:Problem} does not have a solution when $\alpha<\alpha^*$ since the constraint set is empty. Therefore, estimating the value of the minimum sum-rate $\alpha^*$ helps determine whether an $\alpha$ is feasible or not, and, as discussed before, reduce the overall complexity of solving problem \eqref{eq:Problem}.

\section{Minimum Sum-rate}
\label{sec:MinSumRate}

As discussed in Section~\ref{sec:system}, estimating the minimum sum-rate $\alpha^*$ is related to the nonemptyness of $\Ru$. In this section, we show a method to determine $\alpha^*$ by studying the condition for the nonemptyness of $\Ru$. We use a necessary and sufficient condition given in \cite{Fujishige2005} for $\Ru$ to be nonempty to show how to determine $\alpha^*$.

\subsection{Constant Sum-rate Set}

Let $\rv(\X)=\sum_{j\in\X}r_j$ and define
\begin{equation}
g(\X) = \Big| \bigcap_{j\in\C\setminus\X}\Has_j^{c} \Big|.
\end{equation}
It is proved in \cite{CourtIT2014} that a transmission strategy $\rv$ can achieve universal recovery if
\begin{equation} \label{eq:constr}
\rv(\X)\geq g(\X)
\end{equation}
for all $\X$ such that $\X\subset\C$. The interpretation of \eqref{eq:constr} is: the information sent from $\X$ should be complement to that missing in $\C\setminus\X$, i.e., the total number of packets transmitted by the clients in any set $\X$ should be greater or equal to the number of packets that are commonly missing at the remaining clients.

We can describe the constraint set in problem~\eqref{eq:Problem} as the constant sum-rate set
\begin{equation} \label{eq:Ru}
\Ru=\Big\{\rv\in\Real^{K} \colon \rv(\X)\geq g(\X), \forall\X\subset\C,
        \rv(\C)=\alpha \Big\}
\end{equation}
for a PS-CDE system. For an NPS-CDE system, $\rv\in\Real^{K}$ should be replaced with $\rv\in\N$. Therefore, determining minimum sum-rate is equivalent to finding the smallest value of $\alpha$ such that $\Ru$ is nonempty.

\subsection{Determining Minimum Sum-rate}

It is proved in Lemma 1 in \cite{Milo2012} and Lemma 6 in \cite{CourtIT2014} that $g$ is a crossing supermodular function.\footnote{A set function $f\colon 2^{\C}\mapsto \Real$ is supermodular if for all $\X,\Y\subseteq\C$ if $f(\X)+f(\Y) \leq f(\X\cap\Y)+f(\X\cup\Y)$ for all $\X,\Y\subseteq\C$ such that $\X\cap\Y\neq\emptyset$, $\X-\Y\neq\emptyset$, $\Y-\X\neq\emptyset$ and $\X\cup\Y\neq\C$\cite{Fujishige2005}.} Let $\Iset\subseteq\{1,2,\dotsc\}$ with $2\leq|\Iset|\leq{K}$ and denote $\Pat$ a partition of $\C$.\footnote{A partition $\Pat$ of $\C$ satisfies $\X_i\neq\emptyset$, $\X_i\cap\X_j=\emptyset$ and $\cup_{i\in\Iset}\X_i=\C$ for all $i,j\in\Iset$.} We can use the following proposition to determine whether $\Ru$ is empty or not.

\begin{proposition}[nonemptyness of $\Ru$] \label{prop:NEmpt}
$\Ru$ is nonempty if and only if
\begin{align}
\alpha &\geq \sum_{i\in\Iset} g(\X_i),  \label{eq:Alpha1} \\
\alpha &\leq \sum_{i\in\Iset} \Big( \alpha-g(\C\setminus\X_i) \Big) \label{eq:Alpha2}
\end{align}
for all partitions $\Pat$.
\end{proposition}
\begin{IEEEproof}
This is a direct result of Theorem 2.6 in \cite{Fujishige2005}, which derives the necessary and sufficient condition for a crossing supermodular base polyhedron to be nenempty. The crossing supermodular base polyhedron in our case is $\Ru$.
\end{IEEEproof}

Based on Proposition~\ref{prop:NEmpt}, we can determine the minimum sum-rate $\alpha^*$ by the following theorem.

\begin{theorem} \label{theo:MSumRate}
The minimum sum-rate is
\begin{equation} \label{eq:MSumRate1}
\alpha^* = \max \Big\{ \sum_{i\in\Iset}\frac{g(\C\setminus\X_i)}{|\Iset|-1} \colon \text{all partitions } \Pat \Big\}.
\end{equation}
for PS-CDE systems and
\begin{equation} \label{eq:MSumRate2}
\alpha^* = \max \Big\{ \Big\lceil \sum_{i\in\Iset}\frac{g(\C\setminus\X_i)}{|\Iset|-1} \Big\rceil \colon \text{all partitions } \Pat \Big\}
\end{equation}
for NPS-CDE systems, where $\lceil y \rceil$ is the ceiling function which returns the minimum integer value that is no less than $y$.
\end{theorem}
\begin{IEEEproof}
Eq~\eqref{eq:Alpha2} in Proposition~\ref{prop:NEmpt} can be rewritten as
\begin{equation} \label{eq:Alpha3}
\alpha \geq \sum_{i\in\Iset}\frac{g(\C\setminus\X_i)}{|\Iset|-1}.
\end{equation}
But, since $g(\X\cup\Y) \geq g(\X)+g(\Y)$ for all $\emptyset\neq\X,\Y\subset\C$, it can be shown that
\begin{equation} \label{eq:InEqApp}
\sum_{i\in\Iset}\frac{g(\C\setminus\X_i)}{|\Iset|-1}\geq\sum_{i\in\Iset}g(\X_i).\footnotemark
\end{equation}
\footnotetext{Please see the proof in Appendix~\ref{app1}. }
So, $\Ru$ is nonempty if and only if $\alpha \geq \sum_{i\in\Iset}\frac{g(\C\setminus\X_i)}{|\Iset|-1}$ for all partitions. Therefore, for a PS-CDE system, the minimum sum-rate is determined by \eqref{eq:MSumRate1}. For an NPS-CDE system, $\alpha^*$ must be the smallest integer that is greater or equal to $\sum_{i\in\Iset}\frac{g(\C\setminus\X_i)}{|\Iset|-1}$ for all partitions, which can be expressed by \eqref{eq:MSumRate2}.
\end{IEEEproof}

Consider what \eqref{eq:Alpha1} and \eqref{eq:Alpha3} mean in the CDE system. Assume that the clients can form groups, or coalitions, under the condition that any client can only appear in at most one coalition. Then, any form of coalition can be represented by $\Pat$, a partition of $\C$. Based on $\Pat$, consider determining the value of $\alpha$ such that $\Ru$ is nonempty by using \eqref{eq:constr}. On one hand, $\rv(\X_i) \geq g(\X_i)$ for all $i\in\Iset$. Recall that $\alpha=\rv(\C)=\sum_{i\in\Iset}\rv(\X_i)$. We have the condition $\alpha\geq\sum_{i\in\Iset}g(\X_i)$, which is exactly \eqref{eq:Alpha1}. On the other hand, $\rv(\C\setminus\X_i)\geq g(\C\setminus\X_i)$ for all $i\in\Iset$. We have $\sum_{i\in\Iset}\rv(\C\setminus\X_i)\geq\sum_{i\in\Iset}g(\C\setminus\X_i)$. It is equivalent to $(|\Iset|-1)\alpha\geq\sum_{i\in\Iset}g(\C\setminus\X_i)$, which is exactly \eqref{eq:Alpha3}. Since \eqref{eq:constr} should be satisfied for all subsets that are not equal to $\C$, \eqref{eq:Alpha1} and \eqref{eq:Alpha3} should be satisfied for all partitions. But, according to \eqref{eq:InEqApp}, \eqref{eq:Alpha3} is more strict than \eqref{eq:Alpha1}. So, satisfying \eqref{eq:Alpha3} for all partitions is sufficient to determine $\alpha^*$ as stated in Theorem~\ref{theo:MSumRate}.

\begin{example} \label{ex:MinSumRate}
Consider the CDE system in Fig.~\ref{fig:CDESystem}. We have function $g$ as
\begin{align}
&g(\emptyset)=0,g(\{1\})=1,g(\{2\})=0,g(\{3\})=0, \nonumber \\
&g(\{1,2\})=3,g(\{1,3\})=3,g(\{2,3\})=1.   \nonumber
\end{align}
By applying Theorem~\ref{theo:MSumRate}, we have $\alpha^*=3.5$ for PS-CDE system and $\alpha^*=4$ for NPS-CDE system. The corresponding strategy sets are $\R_{3.5}=\{(2.5,0.5,0.5)\}$ and $\R_{4}=\{(2,1,1),(3,0,1),(3,1,0)\}$. It can be shown that $\Ru$ is empty for all $\alpha<3.5$.
\end{example}

\subsection{Relationship with Existing Works}
\label{sec:relation1}

In \cite{MiloDivConq2011}, it was shown that the minimum sum-rate is determined by
\begin{multline} \label{eq:MSumRateChan}
    \alpha^*=L-\min \Big\{ \frac{\sum_{i\in\Iset}|\bigcup_{j\in\X_i}\Has_j|-L}{|\Iset|-1} \colon \\
             \text{all partitions } \Pat \Big\}.\footnotemark
\end{multline}
One can show that it is exactly \eqref{eq:MSumRate1}. Alternatively speaking, the authors in \cite{MiloDivConq2011} derive the the minimum sum-rate for PS-CDE systems.
\footnotetext{It is found in \cite{Court2011} that the solutions of CDE and secrecy generation problems can be found by solving the same linear programming (LP) problem. Eq~\eqref{eq:MSumRateChan} was originally proposed in \cite{Chan2010} to solve a secrecy generation problem. It is also a method to determine the minimum sum-rate of CDE systems \cite{MiloDivConq2011}.}


For PS-CDE systems, Theorem 3 in \cite{CourtIT2014} states that dividing packet into $K-1$ chunks is sufficient to achieve the normalized minimum sum-rate with high probability. But, Theorem~\ref{theo:MSumRate} in this paper establishes that dividing each packet into $K-1$ chunks is sufficient to achieve the normalized minimum sum-rate for sure. In addition, define
\begin{equation} \label{eq:VuPS}
     \hat{g} = (K-1)g(\X).
\end{equation}
We can also conclude from Theorem~\ref{theo:MSumRate} that there exists a minimum integer valued $\alpha$ such that
\begin{equation} \label{eq:RuPS}
\RuPS=\Big\{\rv\in\N \colon \rv(\X)\geq \hat{g}(\X), \forall\X\subset\C,\rv(\C)=\alpha \Big\}
\end{equation}
is nonempty, and that value is the non-normalized minimum sum-rate for the PS-CDE system. Therefore, any PS-CDE problem can be converted to NPS-CDE one (See the example below). For this reason, in the rest of this paper, we consider NPS-CDE problems only.

\begin{example}
Consider the CDE system in Fig.~\ref{fig:CDESystem} and assume it is a PS-CDE system. Let each packet be divided into $K-1=2$ chunks. We replace $g$ with $\hat{g}$ as
\begin{align}
&\hat{g}(\emptyset)=0,\hat{g}(\{1\})=2,\hat{g}(\{2\})=0,\hat{g}(\{3\})=0, \nonumber \\
&\hat{g}(\{1,2\})=6,\hat{g}(\{1,3\})=6,\hat{g}(\{2,3\})=2. \nonumber
\end{align}
By using \eqref{eq:MSumRate2} in Theorem~\ref{theo:MSumRate}, we get $\alpha^*=7$. Therefore, the normalized minimum sum-rate is $7/2=3.5$, which is in consistence with the result in Example~\ref{ex:MinSumRate}.
\end{example}

	\begin{algorithm} [t]
	\label{algo:DSA}
	\small
	\SetAlgoLined
	\SetKwInOut{Input}{input}\SetKwInOut{Output}{output}
	\SetKwFor{For}{for}{do}{endfor}
	\BlankLine
        Initiate the sum-rate lower bound: $\beta=\big\lceil \sum_{j\in\C} \frac{g(\C\setminus\{j\})}{K-1} \big\rceil$\;
        \For {$k=1$ \emph{\KwTo} $K$} {
            $\W_1=\{k\}$\;
            \For {$m=2$ \emph{\KwTo} $K-1$} {
                 $u_m^*=\arg\max \{g(\C\setminus(\W_{m-1}\cup\{u\}))-g(\C\setminus\{u\}) \colon u\in\C\setminus\W_{m-1} \}$\;
                 $\W_m=\W_{m-1}\cup\{u_m^*\}$\;
                 $\beta=\max\{\beta, \big\lceil \frac{g(\C\setminus\W_m)+\sum_{j\in\C\setminus\W_m}g(\C\setminus\{j\})}{|\C\setminus\W_m|} \big\rceil \}$\;
            }
        }
        Output $\beta$\;
	\caption{Deterministic Algorithm}
	\end{algorithm}

\section{Tight Lower Bound on Minimum Sum-rate}
\label{sec:TLBound}

Problem \eqref{eq:MSumRate2} is equivalent to a minimum $k$-partition problem. The NP-hardness of this problem is proved in \cite{Goldschmidt1994}, i.e., to directly determine the minimum sum-rate by Theorem~\ref{theo:MSumRate} is intractable. Therefore, we propose a deterministic algorithm as shown in Algorithm 1 to approximate $\alpha^*$. In this algorithm, we initiate $\beta$, the estimation of the minimum sum-rate, as the value of $\big\lceil \sum_{i\in\Iset}\frac{g(\C\setminus\X_i)}{|\Iset|-1} \big\rceil$ in the case of $K$-partition. We then update $\beta$ by considering the value of $\big\lceil \sum_{i\in\Iset}\frac{g(\C\setminus\X_i)}{|\Iset|-1} \big\rceil$ under a set of $m$-partitions where $2\leq{m}\leq{K-1}$. From the following theorem, we show that the output $\beta$ of Algorithm 1 is a lower bound on $\alpha^*$.

\begin{theorem} \label{theo:TLBound}
For any CDE system, the output $\beta$ of Algorithm 1 satisfies $\beta\leq\alpha^*$.
\end{theorem}
\begin{IEEEproof}
We remark that $\W_m$ in Algorithm 1 for all $2\leq{m}\leq{K-1}$ is generated based on Queyranne's algorithm proposed in \cite{Queyranne}. Here, $\W_m$ and all $j\in\C\setminus{\W_m}$ form a $|\C\setminus{\W_m}|+1$-partition that contains $|\C\setminus{\W_m}|$ singletons. It is proved in \cite{Queyranne} that
\begin{equation}
    g(\C\setminus\W_m) + g(\C\setminus\{j\}) \geq g(\C\setminus(\W_m\setminus\X)) + g(\C\setminus(\X\cup\{j\})),  \nonumber
\end{equation}
for all $j\in\C\setminus\W_m$, $\X\subseteq\W_{m-1}$ and $2\leq{m}\leq{K-1}$. Therefore,
\begin{align}
    &\quad  g(\C\setminus\W_m) + \sum_{j\in\C\setminus{\W_m}} g(\C\setminus\{j\}) \nonumber \\
    &\geq g(\C\setminus(\W_m\setminus\X)) + g(\C\setminus(\{j'\}\cup\X)) \nonumber \\
    &\qquad + \sum_{j\in\C\setminus(\W_m\cup\{j'\})} g(\C\setminus\{j\}),  \nonumber
\end{align}
i.e., $\big\lceil \frac{g(\C\setminus\W_m)+\sum_{j\in\C\setminus\W_m}g(\C\setminus\{j\})}{|\C\setminus\W_m|} \big\rceil$ is the maximum value of $\big\lceil \sum_{i\in\X_i}\frac{g(\C\setminus\X_i)}{|\Iset|-1} \big\rceil$ over a subset of $|\C\setminus{\W_m}|+1$-partitions for all $m$. Therefore, based on Theorem~\ref{theo:MSumRate}, the output $\beta$ of Algorithm 1 is a lower bound on minimum sum-rate $\alpha^*$.
\end{IEEEproof}

\subsection{Relationship with Existing Works}
\label{sec:relation2}

In Lemma 2 in \cite{Roua2010}, it is shown that the minimum sum-rate is lower bounded as
\begin{equation}  \label{eq:LBOld1}
    \alpha^* \geq \max_{j\in\C}|\Has_j^c| = \max_{j\in\C}g(\C\setminus\{j\}).
\end{equation}
In Lemma 10 in \cite{SprintRand2010}, it is shown that the minimum sum-rate is lower bounded as
\begin{equation}  \label{eq:LBOld2}
    \alpha^* \geq \Big\lceil \frac{\sum_{j\in\C}|\Has_j^c|}{K-1} \Big\rceil = \Big\lceil \frac{\sum_{j\in\C}g(\C\setminus\{j\})}{K-1} \Big\rceil.
\end{equation}
But, it can be seen that the output $\beta$ of Algorithm 1 satisfies
\begin{equation}
 \beta \geq \max \Big\{  \max_{j\in\C}\{g(\C\setminus\{j\})+g(\{j\})\},  \Big\lceil \frac{\sum_{j\in\C}g(\C\setminus\{j\})}{K-1} \Big\rceil \Big\}. \nonumber
\end{equation}
Since $\max_{j\in\C}\{g(\C\setminus\{j\})+g(\{j\})\}\geq\max_{j\in\C}g(\C\setminus\{j\})$, $\beta$ is tighter than the lower bounds given in \cite{Roua2010,SprintRand2010}. To see how tight $\beta$ is, we run the following experiments.

\begin{figure}[tbp]
	\centering
    \scalebox{0.7}{
%
%
\begin{tikzpicture}

\begin{axis}[%
width=3.8in,
height=1.4in,
view={-37.5}{30},
scale only axis,
xmin=6,
xmax=30,
xlabel={\large the number of packets $L$},
xmajorgrids,
ymin=3,
ymax=15,
ylabel={\large the number of clients $K$},
ymajorgrids,
zmin=0,
zmax=4,
zmajorgrids,
zlabel={\large average error},
axis x line*=bottom,
axis y line*=left,
axis z line*=left
]

\addplot3[%
surf,
shader=faceted,
draw=black,
colormap/jet,
mesh/rows=25]
table[row sep=crcr,header=false] {
6 3 0.187\\
6 4 0.323\\
6 5 0.423\\
6 6 0.524\\
6 7 0.617\\
6 8 0.686\\
6 9 0.744\\
6 10 0.809\\
6 11 0.862\\
6 12 0.923\\
6 13 0.923\\
6 14 0.983\\
6 15 1.019\\
7 3 0.205\\
7 4 0.326\\
7 5 0.499\\
7 6 0.59\\
7 7 0.747\\
7 8 0.792\\
7 9 0.856\\
7 10 0.88\\
7 11 0.989\\
7 12 1.068\\
7 13 1.101\\
7 14 1.108\\
7 15 1.153\\
8 3 0.208\\
8 4 0.375\\
8 5 0.582\\
8 6 0.678\\
8 7 0.82\\
8 8 0.881\\
8 9 0.956\\
8 10 1.03\\
8 11 1.08\\
8 12 1.133\\
8 13 1.218\\
8 14 1.248\\
8 15 1.259\\
9 3 0.227\\
9 4 0.428\\
9 5 0.581\\
9 6 0.687\\
9 7 0.847\\
9 8 0.969\\
9 9 1.041\\
9 10 1.117\\
9 11 1.209\\
9 12 1.239\\
9 13 1.326\\
9 14 1.377\\
9 15 1.419\\
10 3 0.274\\
10 4 0.426\\
10 5 0.629\\
10 6 0.719\\
10 7 0.879\\
10 8 0.968\\
10 9 1.125\\
10 10 1.201\\
10 11 1.264\\
10 12 1.39\\
10 13 1.37\\
10 14 1.444\\
10 15 1.501\\
11 3 0.236\\
11 4 0.443\\
11 5 0.609\\
11 6 0.802\\
11 7 0.931\\
11 8 1.056\\
11 9 1.129\\
11 10 1.275\\
11 11 1.311\\
11 12 1.39\\
11 13 1.561\\
11 14 1.533\\
11 15 1.626\\
12 3 0.237\\
12 4 0.442\\
12 5 0.632\\
12 6 0.855\\
12 7 0.968\\
12 8 1.133\\
12 9 1.219\\
12 10 1.337\\
12 11 1.423\\
12 12 1.454\\
12 13 1.581\\
12 14 1.599\\
12 15 1.685\\
13 3 0.242\\
13 4 0.448\\
13 5 0.634\\
13 6 0.868\\
13 7 1.025\\
13 8 1.149\\
13 9 1.285\\
13 10 1.349\\
13 11 1.468\\
13 12 1.524\\
13 13 1.67\\
13 14 1.718\\
13 15 1.744\\
14 3 0.263\\
14 4 0.462\\
14 5 0.7\\
14 6 0.844\\
14 7 1.02\\
14 8 1.163\\
14 9 1.304\\
14 10 1.427\\
14 11 1.57\\
14 12 1.652\\
14 13 1.706\\
14 14 1.82\\
14 15 1.848\\
15 3 0.261\\
15 4 0.447\\
15 5 0.686\\
15 6 0.85\\
15 7 1.09\\
15 8 1.201\\
15 9 1.354\\
15 10 1.472\\
15 11 1.607\\
15 12 1.7\\
15 13 1.797\\
15 14 1.852\\
15 15 1.926\\
16 3 0.227\\
16 4 0.45\\
16 5 0.674\\
16 6 0.865\\
16 7 1.078\\
16 8 1.222\\
16 9 1.364\\
16 10 1.556\\
16 11 1.696\\
16 12 1.798\\
16 13 1.824\\
16 14 1.88\\
16 15 1.979\\
17 3 0.23\\
17 4 0.464\\
17 5 0.682\\
17 6 0.897\\
17 7 1.105\\
17 8 1.286\\
17 9 1.462\\
17 10 1.634\\
17 11 1.673\\
17 12 1.855\\
17 13 1.864\\
17 14 2\\
17 15 2.029\\
18 3 0.257\\
18 4 0.418\\
18 5 0.671\\
18 6 0.934\\
18 7 1.113\\
18 8 1.316\\
18 9 1.442\\
18 10 1.575\\
18 11 1.682\\
18 12 1.9\\
18 13 1.947\\
18 14 2.031\\
18 15 2.155\\
19 3 0.192\\
19 4 0.419\\
19 5 0.644\\
19 6 0.904\\
19 7 1.177\\
19 8 1.344\\
19 9 1.542\\
19 10 1.631\\
19 11 1.836\\
19 12 1.912\\
19 13 1.992\\
19 14 2.102\\
19 15 2.114\\
20 3 0.211\\
20 4 0.413\\
20 5 0.671\\
20 6 0.857\\
20 7 1.114\\
20 8 1.321\\
20 9 1.585\\
20 10 1.594\\
20 11 1.809\\
20 12 1.942\\
20 13 2.079\\
20 14 2.135\\
20 15 2.289\\
21 3 0.214\\
21 4 0.417\\
21 5 0.681\\
21 6 0.905\\
21 7 1.181\\
21 8 1.354\\
21 9 1.499\\
21 10 1.697\\
21 11 1.845\\
21 12 1.944\\
21 13 2.103\\
21 14 2.211\\
21 15 2.31\\
22 3 0.229\\
22 4 0.471\\
22 5 0.71\\
22 6 0.92\\
22 7 1.13\\
22 8 1.415\\
22 9 1.557\\
22 10 1.754\\
22 11 1.907\\
22 12 2.103\\
22 13 2.067\\
22 14 2.231\\
22 15 2.405\\
23 3 0.197\\
23 4 0.469\\
23 5 0.7\\
23 6 0.912\\
23 7 1.156\\
23 8 1.367\\
23 9 1.564\\
23 10 1.77\\
23 11 1.937\\
23 12 2.035\\
23 13 2.235\\
23 14 2.34\\
23 15 2.429\\
24 3 0.201\\
24 4 0.39\\
24 5 0.631\\
24 6 0.905\\
24 7 1.19\\
24 8 1.399\\
24 9 1.618\\
24 10 1.697\\
24 11 2.014\\
24 12 2.036\\
24 13 2.241\\
24 14 2.324\\
24 15 2.432\\
25 3 0.166\\
25 4 0.418\\
25 5 0.655\\
25 6 0.877\\
25 7 1.181\\
25 8 1.42\\
25 9 1.651\\
25 10 1.816\\
25 11 1.989\\
25 12 2.124\\
25 13 2.242\\
25 14 2.381\\
25 15 2.45\\
26 3 0.203\\
26 4 0.399\\
26 5 0.618\\
26 6 0.925\\
26 7 1.208\\
26 8 1.39\\
26 9 1.615\\
26 10 1.819\\
26 11 1.98\\
26 12 2.128\\
26 13 2.295\\
26 14 2.415\\
26 15 2.529\\
27 3 0.179\\
27 4 0.415\\
27 5 0.6\\
27 6 0.945\\
27 7 1.168\\
27 8 1.468\\
27 9 1.703\\
27 10 1.885\\
27 11 2.049\\
27 12 2.185\\
27 13 2.325\\
27 14 2.38\\
27 15 2.585\\
28 3 0.171\\
28 4 0.415\\
28 5 0.64\\
28 6 0.908\\
28 7 1.193\\
28 8 1.447\\
28 9 1.67\\
28 10 1.836\\
28 11 1.993\\
28 12 2.21\\
28 13 2.401\\
28 14 2.479\\
28 15 2.612\\
29 3 0.21\\
29 4 0.384\\
29 5 0.613\\
29 6 0.931\\
29 7 1.177\\
29 8 1.398\\
29 9 1.623\\
29 10 1.889\\
29 11 2.096\\
29 12 2.263\\
29 13 2.427\\
29 14 2.48\\
29 15 2.594\\
30 3 0.189\\
30 4 0.388\\
30 5 0.662\\
30 6 0.888\\
30 7 1.158\\
30 8 1.556\\
30 9 1.756\\
30 10 1.992\\
30 11 2.106\\
30 12 2.374\\
30 13 2.548\\
30 14 2.607\\
30 15 2.663\\
};
\end{axis}
\end{tikzpicture}%
}
	\caption{The average error of lower bound in \cite{SprintRand2010} when $K$ varies from $3$ to $15$ and $L$ varies from $6$ to $30$.}
	\label{fig:ErrLBOld2}
\end{figure}
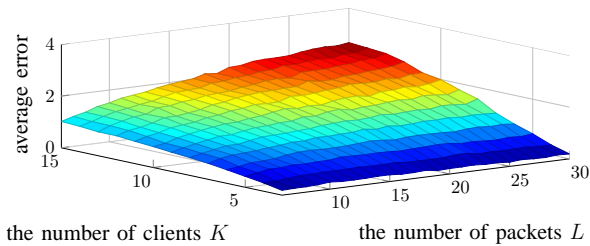

\begin{example} \label{ex:MainEx}
We vary the number of clients $K$ from $3$ to $15$ and the number of packets $L$ from $6$ to $30$. For each combination of $K$ and $L$, we repeat the procedure below for $1000$ times.
\begin{itemize}
    \item randomly generate the has-sets $\Has_j$ for all ${j\in\C}$ subject to the condition $\cup_{j\in\C}\Has_j=\Pak$;
    \item get the minimum sum-rate $\alpha^*$ by using the randomized algorithm proposed in \cite{SprintRand2010}; calculate the lower bounds given in \cite{Roua2010} and \cite{SprintRand2010}; obtain the lower bound $\beta$ by running Algorithm 1.
\end{itemize}
We take the error as the absolute value of the difference between the lower bound and the minimum sum-rate $\alpha^*$ and the average error as the mean error over $1000$ repetitions. We show the average error of lower bounds in \cite{Roua2010}, \cite{SprintRand2010} and Algorithm 1 in Figs.~\ref{fig:ErrLBOld1}, \ref{fig:ErrLBOld2} and \ref{fig:ErrTLB}, respectively. It can be seen that the lower bound found by Algorithm 1 is much tighter than the ones in \cite{Roua2010,SprintRand2010} and, in most cases, Algorithm 1 finds the exact value of $\alpha^*$.
\end{example}

\subsection{Complexity}

It should be noted that running Algorithm 1 alone is not able to completely solve problem~\eqref{eq:Problem}. To find a solution of problem~\eqref{eq:Problem} and also check if the universal recovery can be achieved with the sum-rate $\alpha=\beta$, one needs to run the algorithms proposed in \cite{CourtIT2014,Milo2012}. But, running Algorithm 1 before those algorithms in \cite{CourtIT2014,Milo2012} helps reduce the overall complexity.

\begin{figure}[tbp]
	\centering
    \scalebox{0.7}{
%
%
\begin{tikzpicture}

\begin{axis}[%
width=3.8in,
height=1.4in,
view={-37.5}{30},
scale only axis,
xmin=6,
xmax=30,
xlabel={\large the number of packets $L$},
xmajorgrids,
ymin=3,
ymax=15,
ylabel={\large the number of clients $K$},
ymajorgrids,
zmin=0,
zmax=0.003,
zmajorgrids,
zlabel={\large average error},
axis x line*=bottom,
axis y line*=left,
axis z line*=left
]

\addplot3[%
surf,
shader=faceted,
draw=black,
colormap/jet,
mesh/rows=25]
table[row sep=crcr,header=false] {
6 3 0\\
6 4 0.001\\
6 5 0\\
6 6 0\\
6 7 0\\
6 8 0\\
6 9 0\\
6 10 0\\
6 11 0.001\\
6 12 0\\
6 13 0\\
6 14 0\\
6 15 0\\
7 3 0\\
7 4 0\\
7 5 0\\
7 6 0.003\\
7 7 0\\
7 8 0\\
7 9 0\\
7 10 0\\
7 11 0\\
7 12 0\\
7 13 0\\
7 14 0\\
7 15 0\\
8 3 0\\
8 4 0\\
8 5 0\\
8 6 0\\
8 7 0\\
8 8 0\\
8 9 0\\
8 10 0\\
8 11 0\\
8 12 0\\
8 13 0\\
8 14 0\\
8 15 0\\
9 3 0\\
9 4 0\\
9 5 0\\
9 6 0\\
9 7 0.001\\
9 8 0\\
9 9 0\\
9 10 0\\
9 11 0\\
9 12 0\\
9 13 0\\
9 14 0\\
9 15 0\\
10 3 0\\
10 4 0\\
10 5 0\\
10 6 0\\
10 7 0\\
10 8 0\\
10 9 0\\
10 10 0\\
10 11 0\\
10 12 0\\
10 13 0\\
10 14 0\\
10 15 0\\
11 3 0\\
11 4 0.001\\
11 5 0\\
11 6 0\\
11 7 0\\
11 8 0\\
11 9 0\\
11 10 0\\
11 11 0\\
11 12 0\\
11 13 0\\
11 14 0\\
11 15 0\\
12 3 0\\
12 4 0\\
12 5 0\\
12 6 0\\
12 7 0\\
12 8 0\\
12 9 0\\
12 10 0\\
12 11 0\\
12 12 0\\
12 13 0\\
12 14 0\\
12 15 0\\
13 3 0\\
13 4 0\\
13 5 0\\
13 6 0\\
13 7 0\\
13 8 0\\
13 9 0\\
13 10 0\\
13 11 0\\
13 12 0\\
13 13 0\\
13 14 0\\
13 15 0\\
14 3 0\\
14 4 0\\
14 5 0.001\\
14 6 0\\
14 7 0\\
14 8 0.001\\
14 9 0\\
14 10 0\\
14 11 0\\
14 12 0\\
14 13 0\\
14 14 0\\
14 15 0\\
15 3 0\\
15 4 0\\
15 5 0\\
15 6 0\\
15 7 0\\
15 8 0\\
15 9 0\\
15 10 0\\
15 11 0\\
15 12 0\\
15 13 0\\
15 14 0\\
15 15 0\\
16 3 0\\
16 4 0\\
16 5 0\\
16 6 0.001\\
16 7 0\\
16 8 0\\
16 9 0\\
16 10 0\\
16 11 0\\
16 12 0\\
16 13 0\\
16 14 0\\
16 15 0\\
17 3 0\\
17 4 0\\
17 5 0\\
17 6 0\\
17 7 0\\
17 8 0\\
17 9 0\\
17 10 0\\
17 11 0\\
17 12 0\\
17 13 0\\
17 14 0\\
17 15 0\\
18 3 0\\
18 4 0\\
18 5 0\\
18 6 0\\
18 7 0\\
18 8 0\\
18 9 0\\
18 10 0\\
18 11 0\\
18 12 0\\
18 13 0\\
18 14 0\\
18 15 0\\
19 3 0\\
19 4 0\\
19 5 0\\
19 6 0\\
19 7 0\\
19 8 0\\
19 9 0\\
19 10 0\\
19 11 0\\
19 12 0\\
19 13 0\\
19 14 0\\
19 15 0\\
20 3 0\\
20 4 0\\
20 5 0\\
20 6 0\\
20 7 0\\
20 8 0\\
20 9 0\\
20 10 0\\
20 11 0\\
20 12 0\\
20 13 0\\
20 14 0\\
20 15 0\\
21 3 0\\
21 4 0.001\\
21 5 0\\
21 6 0\\
21 7 0\\
21 8 0\\
21 9 0\\
21 10 0\\
21 11 0\\
21 12 0\\
21 13 0\\
21 14 0\\
21 15 0\\
22 3 0\\
22 4 0\\
22 5 0\\
22 6 0\\
22 7 0\\
22 8 0\\
22 9 0\\
22 10 0\\
22 11 0\\
22 12 0\\
22 13 0\\
22 14 0\\
22 15 0\\
23 3 0\\
23 4 0\\
23 5 0\\
23 6 0\\
23 7 0.001\\
23 8 0\\
23 9 0\\
23 10 0\\
23 11 0\\
23 12 0\\
23 13 0\\
23 14 0\\
23 15 0\\
24 3 0\\
24 4 0\\
24 5 0\\
24 6 0\\
24 7 0\\
24 8 0\\
24 9 0\\
24 10 0\\
24 11 0\\
24 12 0\\
24 13 0\\
24 14 0\\
24 15 0\\
25 3 0\\
25 4 0\\
25 5 0\\
25 6 0\\
25 7 0\\
25 8 0\\
25 9 0\\
25 10 0\\
25 11 0\\
25 12 0\\
25 13 0\\
25 14 0\\
25 15 0\\
26 3 0\\
26 4 0\\
26 5 0\\
26 6 0\\
26 7 0\\
26 8 0\\
26 9 0\\
26 10 0\\
26 11 0\\
26 12 0\\
26 13 0\\
26 14 0\\
26 15 0\\
27 3 0\\
27 4 0\\
27 5 0\\
27 6 0\\
27 7 0\\
27 8 0\\
27 9 0\\
27 10 0\\
27 11 0\\
27 12 0\\
27 13 0\\
27 14 0\\
27 15 0\\
28 3 0\\
28 4 0\\
28 5 0\\
28 6 0\\
28 7 0\\
28 8 0\\
28 9 0\\
28 10 0\\
28 11 0\\
28 12 0\\
28 13 0\\
28 14 0\\
28 15 0\\
29 3 0\\
29 4 0\\
29 5 0\\
29 6 0\\
29 7 0\\
29 8 0\\
29 9 0\\
29 10 0\\
29 11 0\\
29 12 0\\
29 13 0\\
29 14 0\\
29 15 0\\
30 3 0\\
30 4 0\\
30 5 0\\
30 6 0\\
30 7 0\\
30 8 0\\
30 9 0\\
30 10 0\\
30 11 0\\
30 12 0\\
30 13 0\\
30 14 0\\
30 15 0\\
};
\end{axis}
\end{tikzpicture}%
}
	\caption{The average error of lower bound $\beta$ found by Algorithm 1 when $K$ varies from $3$ to $15$ and $L$ varies from $6$ to $30$.}
	\label{fig:ErrTLB}
\end{figure}
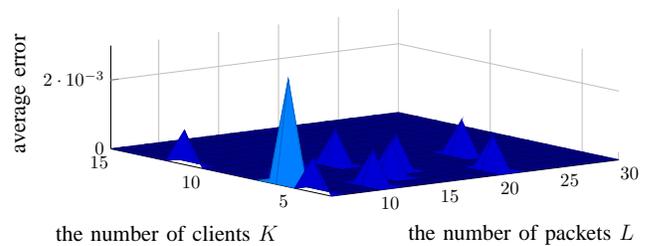

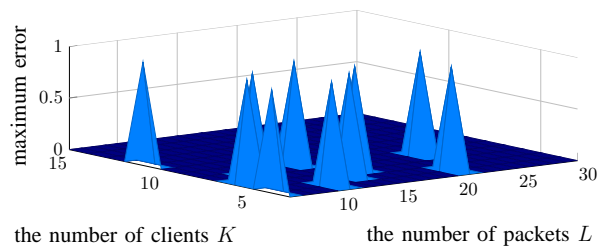
\begin{figure}[tbp]
	\centering
    \scalebox{0.7}{
%
%
\begin{tikzpicture}

\begin{axis}[%
width=3.8in,
height=1.4in,
view={-37.5}{30},
scale only axis,
xmin=6,
xmax=30,
xlabel={\large the number of packets $L$},
xmajorgrids,
ymin=3,
ymax=15,
ylabel={\large the number of clients $K$},
ymajorgrids,
zmin=0,
zmax=1,
zmajorgrids,
zlabel={\large maximum error},
axis x line*=bottom,
axis y line*=left,
axis z line*=left
]

\addplot3[%
surf,
shader=faceted,
draw=black,
colormap/jet,
mesh/rows=25]
table[row sep=crcr,header=false] {
6 3 0\\
6 4 1\\
6 5 0\\
6 6 0\\
6 7 0\\
6 8 0\\
6 9 0\\
6 10 0\\
6 11 1\\
6 12 0\\
6 13 0\\
6 14 0\\
6 15 0\\
7 3 0\\
7 4 0\\
7 5 0\\
7 6 1\\
7 7 0\\
7 8 0\\
7 9 0\\
7 10 0\\
7 11 0\\
7 12 0\\
7 13 0\\
7 14 0\\
7 15 0\\
8 3 0\\
8 4 0\\
8 5 0\\
8 6 0\\
8 7 0\\
8 8 0\\
8 9 0\\
8 10 0\\
8 11 0\\
8 12 0\\
8 13 0\\
8 14 0\\
8 15 0\\
9 3 0\\
9 4 0\\
9 5 0\\
9 6 0\\
9 7 1\\
9 8 0\\
9 9 0\\
9 10 0\\
9 11 0\\
9 12 0\\
9 13 0\\
9 14 0\\
9 15 0\\
10 3 0\\
10 4 0\\
10 5 0\\
10 6 0\\
10 7 0\\
10 8 0\\
10 9 0\\
10 10 0\\
10 11 0\\
10 12 0\\
10 13 0\\
10 14 0\\
10 15 0\\
11 3 0\\
11 4 1\\
11 5 0\\
11 6 0\\
11 7 0\\
11 8 0\\
11 9 0\\
11 10 0\\
11 11 0\\
11 12 0\\
11 13 0\\
11 14 0\\
11 15 0\\
12 3 0\\
12 4 0\\
12 5 0\\
12 6 0\\
12 7 0\\
12 8 0\\
12 9 0\\
12 10 0\\
12 11 0\\
12 12 0\\
12 13 0\\
12 14 0\\
12 15 0\\
13 3 0\\
13 4 0\\
13 5 0\\
13 6 0\\
13 7 0\\
13 8 0\\
13 9 0\\
13 10 0\\
13 11 0\\
13 12 0\\
13 13 0\\
13 14 0\\
13 15 0\\
14 3 0\\
14 4 0\\
14 5 1\\
14 6 0\\
14 7 0\\
14 8 1\\
14 9 0\\
14 10 0\\
14 11 0\\
14 12 0\\
14 13 0\\
14 14 0\\
14 15 0\\
15 3 0\\
15 4 0\\
15 5 0\\
15 6 0\\
15 7 0\\
15 8 0\\
15 9 0\\
15 10 0\\
15 11 0\\
15 12 0\\
15 13 0\\
15 14 0\\
15 15 0\\
16 3 0\\
16 4 0\\
16 5 0\\
16 6 1\\
16 7 0\\
16 8 0\\
16 9 0\\
16 10 0\\
16 11 0\\
16 12 0\\
16 13 0\\
16 14 0\\
16 15 0\\
17 3 0\\
17 4 0\\
17 5 0\\
17 6 0\\
17 7 0\\
17 8 0\\
17 9 0\\
17 10 0\\
17 11 0\\
17 12 0\\
17 13 0\\
17 14 0\\
17 15 0\\
18 3 0\\
18 4 0\\
18 5 0\\
18 6 0\\
18 7 0\\
18 8 0\\
18 9 0\\
18 10 0\\
18 11 0\\
18 12 0\\
18 13 0\\
18 14 0\\
18 15 0\\
19 3 0\\
19 4 0\\
19 5 0\\
19 6 0\\
19 7 0\\
19 8 0\\
19 9 0\\
19 10 0\\
19 11 0\\
19 12 0\\
19 13 0\\
19 14 0\\
19 15 0\\
20 3 0\\
20 4 0\\
20 5 0\\
20 6 0\\
20 7 0\\
20 8 0\\
20 9 0\\
20 10 0\\
20 11 0\\
20 12 0\\
20 13 0\\
20 14 0\\
20 15 0\\
21 3 0\\
21 4 1\\
21 5 0\\
21 6 0\\
21 7 0\\
21 8 0\\
21 9 0\\
21 10 0\\
21 11 0\\
21 12 0\\
21 13 0\\
21 14 0\\
21 15 0\\
22 3 0\\
22 4 0\\
22 5 0\\
22 6 0\\
22 7 0\\
22 8 0\\
22 9 0\\
22 10 0\\
22 11 0\\
22 12 0\\
22 13 0\\
22 14 0\\
22 15 0\\
23 3 0\\
23 4 0\\
23 5 0\\
23 6 0\\
23 7 1\\
23 8 0\\
23 9 0\\
23 10 0\\
23 11 0\\
23 12 0\\
23 13 0\\
23 14 0\\
23 15 0\\
24 3 0\\
24 4 0\\
24 5 0\\
24 6 0\\
24 7 0\\
24 8 0\\
24 9 0\\
24 10 0\\
24 11 0\\
24 12 0\\
24 13 0\\
24 14 0\\
24 15 0\\
25 3 0\\
25 4 0\\
25 5 0\\
25 6 0\\
25 7 0\\
25 8 0\\
25 9 0\\
25 10 0\\
25 11 0\\
25 12 0\\
25 13 0\\
25 14 0\\
25 15 0\\
26 3 0\\
26 4 0\\
26 5 0\\
26 6 0\\
26 7 0\\
26 8 0\\
26 9 0\\
26 10 0\\
26 11 0\\
26 12 0\\
26 13 0\\
26 14 0\\
26 15 0\\
27 3 0\\
27 4 0\\
27 5 0\\
27 6 0\\
27 7 0\\
27 8 0\\
27 9 0\\
27 10 0\\
27 11 0\\
27 12 0\\
27 13 0\\
27 14 0\\
27 15 0\\
28 3 0\\
28 4 0\\
28 5 0\\
28 6 0\\
28 7 0\\
28 8 0\\
28 9 0\\
28 10 0\\
28 11 0\\
28 12 0\\
28 13 0\\
28 14 0\\
28 15 0\\
29 3 0\\
29 4 0\\
29 5 0\\
29 6 0\\
29 7 0\\
29 8 0\\
29 9 0\\
29 10 0\\
29 11 0\\
29 12 0\\
29 13 0\\
29 14 0\\
29 15 0\\
30 3 0\\
30 4 0\\
30 5 0\\
30 6 0\\
30 7 0\\
30 8 0\\
30 9 0\\
30 10 0\\
30 11 0\\
30 12 0\\
30 13 0\\
30 14 0\\
30 15 0\\
};
\end{axis}
\end{tikzpicture}%
}
	\caption{The maximum error of lower bound $\beta$ found by Algorithm 1 when $K$ varies from $3$ to $15$ and $L$ varies from $6$ to $30$.}
	\label{fig:MaxErrTLB}
\end{figure}

As aforementioned, the complexity of algorithms in \cite{CourtIT2014,Milo2012} largely depends on the complexity of solving an submodular function minimization problem. For example, the complexity of the algorithm in \cite{CourtIT2014} is $O(K \cdot SFM(K))$, where $SFM(K)$ is the complexity of running a submodular function minimization algorithm. Although it is proved that submodular function minimization problem can be solved in polynomial time, the lowest complexity of $SFM(K)$ as proposed in \cite{Goemans1995} is still $O(K^5\gamma+K^6)$, where $\gamma$ is the complexity of evaluating function $g$. On the other hand, the algorithms proposed in \cite{CourtIT2014,Milo2012} also require a certain value of $\alpha$ as an input. For example, the authors in \cite{CourtIT2014} suggested use an initial value of $\alpha$ equal to the lower bound in \cite{Roua2010}, repeat running an algorithm with complexity $O(K \cdot SFM(K))$ and increase $\alpha$ until a feasible solution is found. Based on Fig.~\ref{fig:ErrLBOld1}, it means that this algorithm should be called for approximately five times before it finds a feasible solution when $K=3$ and $L=30$. On the contrary, the complexity of of Algorithm 1 is just $O(K^3\gamma)$.\footnote{Algorithm 1 contains $K(K-2)$ iterations, and each iteration involves a maximization over client subset $\C\setminus\W_{m-1}$ by evaluating function $g$. Therefore, the complexity is $O(K^3\gamma)$.} Based on Fig.~\ref{fig:ErrTLB} if we choose the initial value of $\alpha$ to be $\beta$, the number of repetitions of running the algorithm in \cite{CourtIT2014}, and hence the overall complexity, will be largely reduced. In fact, in most cases, the algorithm in \cite{CourtIT2014} will be called just once (since $\beta=\alpha^*$ in most cases). In Fig.~\ref{fig:MaxErrTLB}, we show the maximum error of $\beta$ collected in Example~\ref{ex:MainEx}. It can be seen that the maximum error of $\beta$ is one, which means that if the $\beta\neq\alpha^*$, the number of repetitions of running the algorithm in \cite{CourtIT2014} will not exceed two.

\begin{figure}[tbp]
	\centering
    \scalebox{0.7}{
%
%
\begin{tikzpicture}

\begin{axis}[%
width=3.8in,
height=1.4in,
view={-37.5}{30},
scale only axis,
xmin=6,
xmax=30,
xlabel={\large the number of packets $L$},
xmajorgrids,
ymin=3,
ymax=15,
ylabel={\large the number of clients $K$},
ymajorgrids,
zmin=0,
zmax=6,
zmajorgrids,
zlabel={\large average error},
axis x line*=bottom,
axis y line*=left,
axis z line*=left
]

\addplot3[%
surf,
shader=faceted,
draw=black,
colormap/jet,
mesh/rows=25]
table[row sep=crcr,header=false] {
6 3 0.767\\
6 4 0.404\\
6 5 0.24\\
6 6 0.169\\
6 7 0.108\\
6 8 0.068\\
6 9 0.061\\
6 10 0.049\\
6 11 0.028\\
6 12 0.027\\
6 13 0.021\\
6 14 0.023\\
6 15 0.014\\
7 3 0.961\\
7 4 0.481\\
7 5 0.288\\
7 6 0.159\\
7 7 0.127\\
7 8 0.071\\
7 9 0.052\\
7 10 0.038\\
7 11 0.025\\
7 12 0.036\\
7 13 0.026\\
7 14 0.026\\
7 15 0.021\\
8 3 1.078\\
8 4 0.565\\
8 5 0.33\\
8 6 0.208\\
8 7 0.145\\
8 8 0.098\\
8 9 0.076\\
8 10 0.049\\
8 11 0.042\\
8 12 0.036\\
8 13 0.03\\
8 14 0.022\\
8 15 0.014\\
9 3 1.289\\
9 4 0.665\\
9 5 0.382\\
9 6 0.242\\
9 7 0.163\\
9 8 0.101\\
9 9 0.088\\
9 10 0.055\\
9 11 0.047\\
9 12 0.041\\
9 13 0.025\\
9 14 0.028\\
9 15 0.023\\
10 3 1.42\\
10 4 0.745\\
10 5 0.451\\
10 6 0.286\\
10 7 0.167\\
10 8 0.111\\
10 9 0.089\\
10 10 0.06\\
10 11 0.054\\
10 12 0.046\\
10 13 0.043\\
10 14 0.026\\
10 15 0.025\\
11 3 1.634\\
11 4 0.849\\
11 5 0.487\\
11 6 0.285\\
11 7 0.205\\
11 8 0.14\\
11 9 0.099\\
11 10 0.096\\
11 11 0.07\\
11 12 0.049\\
11 13 0.047\\
11 14 0.026\\
11 15 0.033\\
12 3 1.797\\
12 4 0.907\\
12 5 0.487\\
12 6 0.309\\
12 7 0.207\\
12 8 0.141\\
12 9 0.1\\
12 10 0.091\\
12 11 0.07\\
12 12 0.059\\
12 13 0.04\\
12 14 0.035\\
12 15 0.031\\
13 3 1.917\\
13 4 1.008\\
13 5 0.532\\
13 6 0.337\\
13 7 0.221\\
13 8 0.16\\
13 9 0.127\\
13 10 0.09\\
13 11 0.068\\
13 12 0.049\\
13 13 0.041\\
13 14 0.047\\
13 15 0.046\\
14 3 2.147\\
14 4 1.06\\
14 5 0.634\\
14 6 0.4\\
14 7 0.275\\
14 8 0.185\\
14 9 0.144\\
14 10 0.126\\
14 11 0.085\\
14 12 0.075\\
14 13 0.067\\
14 14 0.052\\
14 15 0.043\\
15 3 2.35\\
15 4 1.241\\
15 5 0.65\\
15 6 0.434\\
15 7 0.304\\
15 8 0.214\\
15 9 0.157\\
15 10 0.11\\
15 11 0.075\\
15 12 0.074\\
15 13 0.057\\
15 14 0.06\\
15 15 0.039\\
16 3 2.5\\
16 4 1.34\\
16 5 0.743\\
16 6 0.464\\
16 7 0.31\\
16 8 0.221\\
16 9 0.174\\
16 10 0.13\\
16 11 0.105\\
16 12 0.091\\
16 13 0.059\\
16 14 0.051\\
16 15 0.034\\
17 3 2.776\\
17 4 1.354\\
17 5 0.802\\
17 6 0.485\\
17 7 0.351\\
17 8 0.244\\
17 9 0.189\\
17 10 0.148\\
17 11 0.092\\
17 12 0.095\\
17 13 0.073\\
17 14 0.048\\
17 15 0.047\\
18 3 2.893\\
18 4 1.508\\
18 5 0.834\\
18 6 0.55\\
18 7 0.372\\
18 8 0.231\\
18 9 0.199\\
18 10 0.137\\
18 11 0.108\\
18 12 0.086\\
18 13 0.075\\
18 14 0.057\\
18 15 0.057\\
19 3 3.031\\
19 4 1.606\\
19 5 0.928\\
19 6 0.554\\
19 7 0.385\\
19 8 0.268\\
19 9 0.186\\
19 10 0.132\\
19 11 0.121\\
19 12 0.102\\
19 13 0.081\\
19 14 0.079\\
19 15 0.047\\
20 3 3.287\\
20 4 1.651\\
20 5 0.931\\
20 6 0.608\\
20 7 0.404\\
20 8 0.314\\
20 9 0.219\\
20 10 0.163\\
20 11 0.138\\
20 12 0.096\\
20 13 0.09\\
20 14 0.079\\
20 15 0.065\\
21 3 3.524\\
21 4 1.78\\
21 5 0.995\\
21 6 0.633\\
21 7 0.371\\
21 8 0.31\\
21 9 0.249\\
21 10 0.177\\
21 11 0.133\\
21 12 0.104\\
21 13 0.095\\
21 14 0.077\\
21 15 0.078\\
22 3 3.717\\
22 4 1.913\\
22 5 1.047\\
22 6 0.641\\
22 7 0.463\\
22 8 0.304\\
22 9 0.217\\
22 10 0.174\\
22 11 0.134\\
22 12 0.112\\
22 13 0.085\\
22 14 0.088\\
22 15 0.089\\
23 3 3.827\\
23 4 2.022\\
23 5 1.115\\
23 6 0.725\\
23 7 0.471\\
23 8 0.347\\
23 9 0.265\\
23 10 0.2\\
23 11 0.17\\
23 12 0.118\\
23 13 0.111\\
23 14 0.1\\
23 15 0.065\\
24 3 4.017\\
24 4 2.096\\
24 5 1.231\\
24 6 0.751\\
24 7 0.478\\
24 8 0.374\\
24 9 0.282\\
24 10 0.202\\
24 11 0.152\\
24 12 0.12\\
24 13 0.109\\
24 14 0.085\\
24 15 0.07\\
25 3 4.2\\
25 4 2.251\\
25 5 1.325\\
25 6 0.797\\
25 7 0.504\\
25 8 0.38\\
25 9 0.302\\
25 10 0.222\\
25 11 0.15\\
25 12 0.139\\
25 13 0.103\\
25 14 0.099\\
25 15 0.081\\
26 3 4.424\\
26 4 2.257\\
26 5 1.323\\
26 6 0.867\\
26 7 0.531\\
26 8 0.365\\
26 9 0.291\\
26 10 0.233\\
26 11 0.18\\
26 12 0.148\\
26 13 0.112\\
26 14 0.089\\
26 15 0.085\\
27 3 4.723\\
27 4 2.411\\
27 5 1.383\\
27 6 0.819\\
27 7 0.584\\
27 8 0.427\\
27 9 0.269\\
27 10 0.216\\
27 11 0.189\\
27 12 0.138\\
27 13 0.114\\
27 14 0.103\\
27 15 0.096\\
28 3 4.819\\
28 4 2.471\\
28 5 1.466\\
28 6 0.915\\
28 7 0.608\\
28 8 0.422\\
28 9 0.299\\
28 10 0.255\\
28 11 0.178\\
28 12 0.15\\
28 13 0.143\\
28 14 0.129\\
28 15 0.092\\
29 3 5.083\\
29 4 2.617\\
29 5 1.556\\
29 6 0.936\\
29 7 0.623\\
29 8 0.428\\
29 9 0.294\\
29 10 0.272\\
29 11 0.187\\
29 12 0.159\\
29 13 0.119\\
29 14 0.128\\
29 15 0.104\\
30 3 5.201\\
30 4 2.776\\
30 5 1.541\\
30 6 0.99\\
30 7 0.683\\
30 8 0.461\\
30 9 0.338\\
30 10 0.276\\
30 11 0.216\\
30 12 0.158\\
30 13 0.151\\
30 14 0.117\\
30 15 0.097\\
};
\end{axis}
\end{tikzpicture}%
}
	\caption{The average error of lower bound in \cite{Roua2010} when $K$ varies from $3$ to $15$ and $L$ varies from $6$ to $30$.}
	\label{fig:ErrLBOld1}
\end{figure}
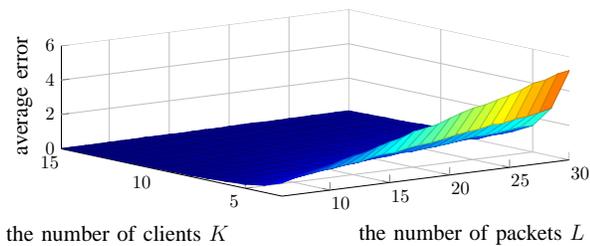

\section{Conclusion}

In this paper, we showed that the minimum sum-rate $\alpha^*$ in a CDE system could be determined by a maximization over all possible partitions of the client set. Instead of solving the maximization problem directly, we proposed a deterministic algorithm to estimate the lower bound on $\alpha^*$ and showed by experiment that this lower bound was much tighter that those derived in the existing literature. The experiment results also showed that the output of Algorithm 1 was the exact value of $\alpha^*$ in most cases and the maximum difference between it and $\alpha^*$ is one, which could avoid repetitively running the existing algorithms for solving CDE problems so that the overall complexity could be reduced.

\section{Acknowledgement}
This work is supported under Australian Research Council Discovery Projects funding scheme (project no. DP120100160).

\bibliographystyle{ieeetr}
\bibliography{CDEMinSumRateBIB}

\appendices

\begin{center}
\Large Appendices
\end{center}

\section{}
\label{app1}
Let $\emptyset\neq\X,\Y\subset\C$ such that $\X\cap\Y=\emptyset$. We have
\begin{align}
    &\quad\ g(\X\cup\Y)-g(\X)-g(\Y)  \nonumber \\
    &=\Big| \bigcap_{j\in\C\setminus(\X\cup\Y)} \Has_j^c \Big| - \Big| \bigcap_{j\in\C\setminus\X} \Has_j^c \Big| - \Big| \bigcap_{j\in\C\setminus\Y} \Has_j^c \Big| \nonumber \\
    &=\Big| \bigcap_{j\in\C\setminus(\X\cup\Y)} \Has_j^c \Big| - \Big| \bigcap_{j\in\C\setminus\X} \Has_j^c  \cup \bigcap_{j\in\C\setminus\Y} \Has_j^c \Big|  \nonumber\\
    &=\Big| \bigcap_{j\in\C\setminus(\X\cup\Y)} \Has_j^c \Big| - \Big| \bigcap_{j\in\C\setminus(\X\cup\Y)} \Has_j^c \cap \Big( \bigcap_{j\in\X} \Has_j^c  \cup \bigcap_{j\in\Y} \Has_j^c \Big) \Big|  \nonumber\\
    &\geq{0}, \nonumber
\end{align}
i.e., $g(\X\cup\Y) \geq g(\X)+g(\Y)$.\footnote{Since $g(\emptyset)=0$, $g$ can be considered supermodular in all $\X$ such that $\emptyset\neq\X\subset\C$.} Then, for any partition $\Pat$ of $\C$,
\begin{equation}
    g(\C\setminus\X_i)\geq\sum_{i'\in\Iset,i'\neq{i}} g(\X_{i'}).
\end{equation}
So, $\sum_{i\in\Iset}g(\C\setminus\X_i) \geq (|\Iset|-1) \sum_{i\in\Iset} g(\X_i)$, i.e.,
\begin{equation}
\sum_{i\in\Iset}\frac{g(\C\setminus\X_i)}{|\Iset|-1}\geq\sum_{i\in\Iset}g(\X_i).
\end{equation}

\end{document}